\documentclass[english, copyright]{eptcs}
\providecommand{\event}{T. Neary, D. Woods, A.K. Seda and N. Murphy (Eds.):\\ The Complexity of Simple Programs 2008.}
\providecommand{\volume}{1}
\providecommand{\anno}{2009}
\providecommand{\firstpage}{123}
\usepackage{breakurl}

\usepackage{amssymb}
\usepackage[latin1]{inputenc}
\usepackage[T1]{fontenc}
\usepackage{babel,graphicx}
\usepackage{amsmath}
\usepackage{amsthm}
\usepackage{stmaryrd}

\newbox\gnBoxA
\newdimen\gnCornerHgt
\setbox\gnBoxA=\hbox{$\ulcorner$}
\global\gnCornerHgt=\ht\gnBoxA
\newdimen\gnArgHgt

\def\Godelnum #1{%
	\setbox\gnBoxA=\hbox{$#1$}%
	\gnArgHgt=\ht\gnBoxA%
	\ifnum \gnArgHgt<\gnCornerHgt
		\gnArgHgt=0pt%
	\else
		\advance \gnArgHgt by -\gnCornerHgt%
	\fi
	\raise\gnArgHgt\hbox{$\ulcorner$} \box\gnBoxA %
		\raise\gnArgHgt\hbox{$\urcorner$}}

\newtheorem{thm}{Theorem}[section]
\newtheorem{cor}[thm]{Corollary}

\DeclareMathOperator{\Prv}{Prov}
\DeclareMathOperator{\Cons}{Cons}

\def\Succ{\varsigma}
\def\Lim{\vartheta}

\title{Busy beavers gone wild}
\author{Gr{\' e}gory Lafitte 
\institute{Laboratoire d'informatique fondamentale de Marseille, CNRS,\\
Aix-Marseille Université, 39, rue F. Joliot-Curie,\\
13453 Marseille Cedex 13, France
\thanks{We acknowledge the support of the French National Research Agency Sycomore grant.}}
\email{Gregory.Lafitte@lif.univ-mrs.fr}
}
\def\titlerunning{Busy beavers gone wild}
\def\authorrunning{G. Lafitte}

\begin{document}
\maketitle

\begin{abstract}
We show some incompleteness results {\em à la\/} Chaitin using the busy beaver functions. Then, with the help of ordinal logics, we show how to obtain a theory in which the values of the busy beaver functions can be provably established and use this to reveal a structure on the provability of the values of these functions.
\end{abstract}

	

\section{Introduction}

When considering a class of Turing machines that halt when started from a blank tape,
{\em busy beavers} for that class, as coined by Tibor Rad\'o, are those Turing machines which eventually halt after the maximum number of steps (or after producing the maximum number of non-blank symbols). Finding for a class of Turing machines the busy beaver champion (for any of the two competitions) is of course uncomputable \cite{Rado62}.
Nevertheless, efforts have been made to try to find the champions starting with the smallest classes of Turing machines. Finding all the different kinds of behavior of Turing machines in a class is quite exhilarating an endeavor.
Rona J. Kopp \cite{Kop81} and Allen H. Brady \cite{Brad83} have thoroughly studied the four states and two symbols class and Heiner Marxen and others \cite{MBun90,MSto90} have tried for more than two decades to experimentally find the busy beaver champion for the five states and two symbols class. 
Heiner Marxen found a candidate but no one has been able to prove that no other machine {\em does better}.
This is most certainly because the behaviors involved are too complicated. 
This state of affairs is an encouragement to the study of what we have called {\em low-level reverse computability}: What does it take to be able to comprehend the behavior of a certain class of Turing machines? 
What is needed to be able to prove these behaviors? These questions raise the links between the study of busy beavers and incompleteness.

\medskip
To start off, there is a need to see the strong links between computability and incompleteness.
For instance, one can obtain a first form of the first incompleteness theorem by considering
propositions of the form $n \not\in X$, where $X$ is a non-recursive but recursively enumerable set, {\it e.g.}, the {\em diagonal halting} set $\mathcal{K}$. Even if the language of the considered theory does not contain $\in$, there is a simple algorithm that generates given $n$ the proposition ``$n \not\in X$''. 
Given a sound (every provable statement is true) recursively enumerable theory $T$, there is a number $n_0$
such that $n_0 \not\in X$ but $T$ does not prove it. The proof is direct: 
Suppose that there is no such $n_0$, then we would have that $T$ proves ``$n \not\in X$'' if and only $n\not \in X$, and $X$ would be recursive (generate the theorems of $T$ and at the same time enumerate $X$; if $n \in X$ then $n$ will eventually show up in the enumeration; otherwise, ``$n\not\in X$'' will eventually show up in the theorems of $T$ and be true by the soundness assumption).
We thus have a true sentence, ``$n_0 \not\in X$'', which is not provable in $T$.

\medskip
Incompleteness is also famously linked to computability via Chaitin's incompleteness theorem.\linebreak
Chaitin's result, showing that there are unprovable statements on Kolmogorov-Chaitin complexity\footnote{\label{Kphi}Loosely speaking, the Kolmogorov-Chaitin complexity of a natural number $n$, denoted by $K(n)$, is the smallest size of a program which generates $n$. 

Another useful variant of Kolmogorov-Chaitin complexity can be defined as follows. To each enumeration $\{\varphi_i\}_{i \in \mathbb{N}}$ of computable partial functions, we associate a variant of $K$: $K_{\varphi}(x | y) = \text{ smallest $e$ such that $\varphi_e (y) = x$, }$ and $K_{\varphi}(x) = K_{\varphi}(x|0)$.

Chaitin's version of Gödel incompleteness is the following theorem.

\begin{thm}[Chaitin's incompleteness]
Let $T$ be an arithmetical-able consistent theory. There is a constant $\mathfrak{c}_T$ such that for all $x$, ``$K(x)>\mathfrak{c}_T$'' is unprovable in $T$.		
\end{thm}

For more on Kolmogorov-Chaitin complexity, the reader is referred to \cite{LV97}. For a formal definition of {\em arithmetical-ability}, see \cite{L-08}.}, is a form of Gödel's first incompleteness theorem. Actually, Kolmogorov showed in the sixties that the set of random (or incompressible) numbers, {\it i.e.}, $\{ x : K(x) \geqslant x \}$, is recursively enumerable but not recursive, and, by the above argument, this is already a version of Gödel's first incompleteness theorem.
Moreover, Kolmogorov's proof can be seen as an application of Berry's paradox (``the least integer not nameable in fewer than seventy characters'' has just now been named in sixty-three characters). Following Boolos \cite{Boo89}, it is thus no wonder that we can get proofs using this Kolmogorov complexity function (or other {\em similar} computability-related functions) of both incompleteness theorems (see \cite{L-08}).

\medskip
One of the reasons of the existence of the quest for better understanding the incompleteness phenomenon lies in the peculiarity of Gödel's unprovable statements. They are not natural mathematical statements: no mathematician has ever stumbled on them (or should we say {\it over them\,}?).
And thus, it seems to many that normal mathematical practice is not concerned with the incompleteness phenomenon.
More and more results show however the contrary. In particular, Harvey Friedman's $\Pi^0_1$ statements, that are unprovable in Zermelo-Fraenkel ({\sc ZF}) set theory and need the $1$-consistency of strong set-theoretical unprovable statements (going way beyond {\sc ZF}) to be proved, are examples of such results.

\medskip
Nevertheless, incompleteness theorems only provide unprovable statements like the {\em consistency} of a theory, that are of an unclear nature. What combinatorial properties does the consistency statement bring to a theory? 
Having a link between consistency (or soundness) and computability, in particular Kolmogorov complexity, would make possible an understanding of what properties consistency adds to a theory. Adding consistency as an axiom would then yield new combinatorial properties because of the existing links between combinatorics and Kolmogorov complexity. 
For a detailed exposition of the links between Gödel incompleteness and Kolmogorov complexity, see \cite{L-08}.

\medskip
In the first part of this article, we start by giving an incompleteness theorem {\em à la\/} Chaitin using the busy beaver functions and then make precise the kind of incompleteness theorems one can obtain with them.  In the second part, we use ordinal logics to construct a theory in which the values of the busy beaver functions can be provably established and use this to find a structure on the provability of the values of these functions.

\section*{Notational conventions}

On top of the usual logical connectives ($\wedge$, $\vee$ and $\neg$), we will respectively denote the logical connectives of {\em implication} and {\em equivalence} by $\rightslice$ and $\equiv$.

\medskip
When we have a recursively enumerable axiomatic theory $T$, $\Cons_T$ designates the arithmetical sentence that there is no proof of $0=1$. When the axioms of $T$ are defined by a formula $\phi$, $\Cons (\phi)$ will designate $\Cons_T$ using that formalization. The provability predicate is designated by $\Prv_T$.

\medskip
The partial recursive functions computed by Turing machines, following a fixed convention, are denoted by $\{\varphi_i\}_{i\in\mathbb{N}}$ (agreeing with the Turing machines' coding: $\varphi_i$ is the partial function {\em computed} by $T_i$).  
The sets $\{W_i\}_{i\in\mathbb{N}}$ denote the recursively enumerable sets, {\it i.e.}, the domains of partial recursive functions.

\medskip
Concerning computability, the reader is referred to \cite{Odif89, Odif99, Roge67, Roge58, Smu93, VS-03}.

\section{Incompleteness and busy-beaverhood}

In the sixties, Tibor Rad\'o, a professor at the Ohio
State University, thought of a simple non-computable function other than
the standard halting problem for Turing machines. Given a fixed
finite number of symbols and states, select those Turing machines which eventually halt when run with a blank tape. Among
these programs, find the maximum number of non-blank symbols left on
the tape when they halt. Alternatively, find the maximum number of
time steps before halting. These functions are well-defined but uncomputable. 
Tibor Rad\'o called them the busy beaver functions. For more on the busy beaver problem, read \cite{Rado62, Lin63, LRad65, Brad66, Kop81, Brad83, Dewd84, Dewd85, Brad88, MSto90, MBun90, LP-07}.

\medskip
Alternative functions can be defined that are close in nature to these busy beaver functions.
Let $\sigma^{\text{steps}}$ be the function which to $i$ gives the maximum number of steps for which a Turing machine with code $\leqslant i$ will keep running before halting, starting with a blank tape. For a Turing machine $M$, $t_M$ denotes the time complexity function of $M$: $t_M (x) = s$ if $M(x)$ halts after $s$ steps.
Following the busy beaver functions' definitions, we define $\sigma^{\text{value}}$ to be the function which to $i$ gives the maximum number which a Turing machine with code $\leqslant i$ will output, following a fixed convention, after halting starting with an input $\leqslant i$.
These functions are in a sense inverses of the $K_\varphi$ function (see footnote \ref{Kphi}). 

\medskip
Other functions can be defined following classical Kolmogorov complexity, {\it e.g.}, the function which to $n$ gives the biggest number with Kolmogorov complexity lower than $n$.

\medskip
We call these functions the $\sigma$ functions.
For each variant, we can define a function focusing on maximizing the {\em number of steps}, {\it e.g.}, $\sigma^{\text{steps}}$, or the {\em output values}, {\it e.g.}, $\sigma^{\text{value}}$. The value of either of these functions on a certain $x$ is computable from $x$ and the value of the other function on input $x+c$ for a certain constant $c$ (see \cite{ShenXX}). 

\medskip
A result similar to Chaitin's incompleteness result concerning Kolmogorov complexity can be obtained concerning the $\sigma$ functions\footnote{A $\sigma$ function is any of the busy beaver functions defined above.}:

\medskip
\begin{thm}[Chaitin-like incompleteness theorem for $\sigma$ functions]
\label{Chaitinlike}
Let $\sigma$ be one of the $\sigma$ functions.
Let $T$ be an arithmetical-able\footnote{For a formal definition of {\em arithmetical-ability}, see \cite{L-08}.} consistent theory. There is a constant $\mathfrak{n}^{\sigma}_T$ such
that 
\begin{equation}
\label{sigma1st}
T \vdash \Cons_T \rightslice \ \forall s \neg \Prv_T ( \Godelnum{\sigma(\mathfrak{n}^{\sigma}_T) < s} ).
\end{equation}
\end{thm}

\proof
Consider a $\Pi_1$ formula $\phi_\sigma$ in the language of $T$ such that $\phi_\sigma (x,s)$ expresses that $\sigma (x) < s$.

Working in $T$, for a given $x$, take the smallest $s$ such that $\Prv_T (\Godelnum{\phi_\sigma (x,s)})$ holds. $T$ being consistent and $\phi_\sigma$ $\Pi_1$,
$\phi_\sigma (x,s)$ also holds.  

$\Prv_T (\Godelnum{\phi_\sigma (x,s)})$ is a $\Sigma_1$ formula and thus can be seen as $\exists y \psi (x,s,y)$ or equivalently $\exists \langle s, y\rangle \psi (x,s,y)$ where $\psi$ is $\Delta_0$. 

Thus there is a Turing machine computing $\psi$. Consider its code $i_\psi$ (or its number of states or transitions, depending on the choice of $\sigma$). For large enough $x$, {\it i.e.}, $x > i_\psi+c$, knowing that $\phi_\sigma (x,s)$ holds (using the computation through shifting, {\it i.e.}, the constant $c$, between both types of $\sigma$ functions), we know that $\sigma (x) < s$ and thus there is an $s' = \langle s'_1, s'_2 \rangle < s$ such that $\psi (x, s'_1, s'_2)$ holds. But for each $s' = \langle s'_1, s'_2 \rangle$ smaller than $s$, the statement $\neg\psi (x,s'_1,s'_2)$ is true by the minimality of $s$, and provable (being $\Delta_0$). Thus we have $\neg \Cons_T$.
\qed

\medskip
We say that a statement $\phi$ is a {\em revelation for $T$} if $\phi$ is unprovable in $T$ and its consistency relative to $T$ (if $T$ is consistent, so is $T+\phi$) is provable from itself in $T$:
 $$T \vdash \phi \rightslice \Cons_T (\phi)$$

\medskip
We have the following result showing how being able to prove an upper bound for a $\sigma$ function can be a revelation.

\medskip
\begin{thm}[Serendipitous incompleteness theorem for $\sigma$ functions] 
\label{serendip}
Let $\sigma$ be one of the $\sigma$ functions.
If $T$ is consistent, then there exists a natural number $\mathfrak{r}^{\sigma}_T$ such that for all $x$, $\sigma (\mathfrak{r}^{\sigma}_T) < x$ is a revelation for~$T$.
\end{thm}

\proof 

Consider the $\Pi_1$ statement $\forall x \ \psi (x)_x$ equivalent to $\Cons_{T+\phi}$.

$\psi \in \Delta_0$ and thus there is a machine $M_\psi$ with code $i_\psi$ such that $M_\psi$ decides $\{x : \psi (x)_x\}$: $M_\psi$ on input $x$ eventually enters an acceptance state if $\psi (x)_x$, or a rejection state otherwise.

Consider another Turing machine $M_\psi'$ which runs $M_\psi$ successively on each natural number starting from $0$ and stops and writes the counter example of $\psi$ if the simulation of $M_\psi$ enters a rejection state.

Let $i_\psi'$ be the code of Turing machine $M_\psi'$. 
$\sigma (i_\psi')$ makes the verification of $\forall x \ \psi (x)_x$ a $\Delta_0$ property.

By using Kleene's recursion theorem on this previous construction, we find $\mathfrak{r}^{\sigma}_T$ such that knowing (or bounding) the value of $\sigma (\mathfrak{r}^{\sigma}_T)$ makes the verification of $\Cons_{T+\sigma (\mathfrak{r}^{\sigma}_T) \leqslant x}$ a $\Delta_0$ property. Knowing that $T$ is consistent and assuming $\sigma (\mathfrak{r}^{\sigma}_T) \leqslant x$, $T$ thus proves $\Cons_T (\sigma (\mathfrak{r}^{\sigma}_T) \leqslant x)$. 

By Gödel's second incompleteness theorem, $\sigma (\mathfrak{r}^{\sigma}_T) \leqslant x$ is an unprovable statement in $T$.
\qed

\medskip
This can be done for other uncomputable functions, like Kolmogorov complexity functions (using the Turing-completeness of their graphs). Truth-table completeness is a plus to be able to have a finite revelation (finite number of constants $\mathfrak{c}^K_{i,T}$ giving the revelation $\bigwedge K(x)>\mathfrak{c}^K_{i,T}$).

\medskip
The same idea can be used to find a constant ${}_{T'}\mathfrak{r}^{\sigma}_T$ such that if $T$ and $T'$ are consistent, then for all $x$, $T \vdash \sigma ({}_{T'}\mathfrak{r}^{\sigma}_T) < x \rightslice \text{$1$-}\Cons_{T'}$. This can be done for other soundness properties than $1$-consistency.

\medskip
By combining the proofs of the two previous theorems \ref{Chaitinlike} and \ref{serendip}, we obtain~: 

\medskip
\begin{cor}[Busy beaver pairs]
\label{bbpairs}
Let $\sigma$ be one of the $\sigma$ functions.
If $T$ is consistent, then there exist $\mathfrak{s}^{\sigma,T}_0$ and $\mathfrak{s}^{\sigma,T}_1$ such that for all $x$, 
$$T \vdash \sigma (\mathfrak{s}^{\sigma,T}_0) < x \rightslice \forall s \neg \Prv_T ( \Godelnum{\sigma(\mathfrak{s}^{\sigma,T}_1) < s} ).$$
\end{cor}

\section{Having enough power to comprehend busy beavers}

We start by recalling the basic notions encompassing ordinal logics. We first need the notion of {\em ordinal notation}.

Let $\langle \mathbf{O}\subset\mathbb{N}, \prec_{\mathbf{O}}, |\cdot| \rangle$ be defined inductively, simultaneously on $\prec_{\mathbf{O}}$ and $|\cdot|:\mathbf{O}\to \text{Ord}$, as follows:

$0\in \mathbf{O}$ and $|0|=0$;

If $a \in \mathbf{O}$, then $\Succ(a) = \langle a, 0\rangle \in \mathbf{O}$, $a \prec_{\mathbf{O}} \Succ(a)$, and $|\Succ(a)| = |a|+1$;

If $\varphi_e$ is a total function which is increasing according to $\prec_{\mathbf{O}}$ (for all $n$, $\varphi_e (n) \prec_{\mathbf{O}} \varphi_e (n+1)$), then $\Lim(e) = \langle e, 1\rangle \in \mathbf{O}$, $\Lim(e) \succ_{\mathbf{O}} \varphi_e(n)$ for all $n$, and $|\Lim(e)| = \lim_{n\in \omega} |\varphi_e (n)|$.

Elements of $\mathbf{O}$ are called {\slshape notations}.
$\Succ(a)$ denotes the {\slshape successor} of $a$, $\Lim(e)$ the {\slshape limit} of $e$.
$\mathbf{O}$ is called the {\itshape general ordinal representation system} because every {\itshape ordinal representation system} is isomorphic to $\mathbf{O}_a = \{b \prec_{\mathbf{O}} a\}$ for a particular $a$.
Because of the definition of $\Lim(e)$, assuming $\varphi_e$ to be total and increasing, $\prec_{\mathbf{O}}$ is uncomputable, even not recursively enumerable. 

\medskip
Starting from a theory $T$, we can now define {\slshape progressions} of theories indexed by notations.

$$
\begin{array}{l}
	T_0 = T\\
	T_{\Succ(a)} = T + \Cons (T)\\
	T_{\Lim(e)} = \bigcup_n T_{\varphi_e (n)}
\end{array}
$$

\medskip
The idea is to have a sequence of theories indexed by ordinals showing {\em how many times} we assume the consistency of the previous theories, the predecessor theory if the ordinal is a successor and all previous theories if a limit ordinal. The difficulty lies in always having recursively enumerable axiomatic theories, {\it i.e.}, having a sequence of formulae $\phi_a$ defining the axioms of $T_a$ and definable from previous $\phi$'s.
Turing \cite{Tur39} and later Feferman \cite{Fef62} came up with this idea of recursive progressions of theories.
Kleene's second recursion theorem shows the existence of such recursive progressions.
Notice that assuming the soundness of the base theory $T$, if we add its consistency to the axioms of $T$, then we know that this enhanced theory is also sound. Hence, soundness is inherited by consistency extensions.

\medskip
Now, the problem is that $T_a$ is not consistent for all $a$ even when assuming the soundness of $T$: by Kleene's recursion theorem, there is $e$ such that $\varphi_e (0) = \Succ(\Lim(e))$ and for such $e$, $T_{\Lim(e)}$ proves all that $T_{\Succ(\Lim(e))}$ proves, in particular $\Cons_{T_{\Lim(e)}}$; thus $T_{\Lim(e)}$ is inconsistent.

\medskip
To ensure that $T_a$ is a consistency extension of $T$, one has to make sure that $a$ belongs to $\mathbf{O}$. One way of doing that is working only with $a$'s from a {\em branch} of $\mathbf{O}$, that is a linearly ordered subset of $\mathbf{O}$ closed under $\prec_{\mathbf{O}}$. A better way is considering only notations, elements of ${\mathfrak{O}}$, such that if $a\in \mathfrak{O}$ and an arithmetically definable variant of ``$b \in \mathbf{O}$'' is provable in $T_a$ then $b\in\mathfrak{O}$.

\medskip
In theorem \ref{pointwisesigma} we use arguments along the lines of Turing's \cite{Tur39} as they were described by Feferman \cite{Fef62, Fef88}.

\medskip
\begin{thm}[Point-wise provability of $\sigma$ values] 
\label{pointwisesigma}
Let $\sigma$ be one of the $\sigma$ functions.
For any consistency progression, for all $a\in\mathbf{O}$ and for all $x\in\mathbb{N}$, there exists $a_x \in \mathbf{O}$ with $a \prec_{\mathbf{O}} a_x$ such that $|a_x|=|a|+\omega+1$ and $T_{a_x}$ proves all true statements of the form ``$\sigma(x)<\cdots$''.
\end{thm}

\proof
Consider a $\Pi_1$ formula $\phi_\sigma (x,s) = \forall t \ \psi_\sigma (t,x,s) $ in the language of $T$ such that $\phi_\sigma (x,s)$ expresses $\sigma (x) < s$ and $\psi_\sigma$ is $\Delta_0$.

Let $\sigma_x$ be the true value of $\sigma(x) +1$.
By Kleene's recursion theorem, let $e_x$ be such that provably in {\scshape PA}, for every $n$,
$$
\varphi_{e_x} (n) = \left\{\begin{array}{cl}
\Succ^n (a) & \text{ if ``$\psi(\Godelnum{i},\Godelnum{x},\Godelnum{\sigma_x})$'' is true for every $i \leq n$, }\\
\Succ(\Lim({e_x})) & \text{ otherwise.}
\end{array}\right.
$$
Since $\forall t \ \psi(t,x,s)$ is true, $\Lim({e_x}) \in \mathbf{O}$ and $|\Lim({e_x})| = |a|+\omega$. 
Let $a_x = \Succ(\Lim({e_x}))$. 
If $\neg \forall t \ \psi(t,x,s)$, then $T_{\varphi_{e_x} (n)}$ is $T_a$ for all sufficiently large $n$ and so $T_{\Lim({e_x})}$ also proves the consistency of $T_{\Lim({e_x})}$. Thus by Gödel's second incompleteness theorem, $T_{\Lim({e_x})}$ is inconsistent.
Hence, since we can prove in $T_{a_x}$ the consistency of $T_{\Lim({e_x})}$, $\forall t\ \psi(t,x,s)$ is also provable in $T_{a_x}$.
\qed

\medskip
For any class of machines, it is thus possible with a combination of theories from a progression of theories to comprehend the busy beaver functions on that class.

\medskip
By an extensive use of Kleene's recursion theorem and by varying the theory in the use of corollary \ref{bbpairs}, we get the following corollary.

\medskip
\begin{cor}[Busy beaver relationships]
Let $\sigma$ be one of the $\sigma$ functions.
If $T$ is consistent, then for all $n>1$, there exist $\{\mathfrak{s}^{\sigma,T}_i\}_{i<n}$ such that for all $s$, $\sigma(\mathfrak{s}^{\sigma,T}_0) < s$ is not provable in $T$, and for all $x$ and $i<(n-1)$, $$T \vdash \sigma (\mathfrak{s}^{\sigma,T}_i) < x \rightslice \forall s \neg \Prv_T ( \Godelnum{\sigma(\mathfrak{s}^{\sigma,T}_{i+1}) < s} ).$$
\end{cor}


\bibliographystyle{eptcs} 

\end{document}